\documentstyle[preprint,aps]{revtex}
\tighten
\begin{document}
\title{Necessary Conditions for Apparent Horizons and
Singularities in Spherically Symmetric Initial Data}
\author{
   Jemal Guven${}^{(1)}$ \thanks{\tt{ jemal@nuclecu.unam.mx}}
and
   Niall \'O Murchadha${}^{(2)}$ \thanks{\tt{ niall@ucc.ie}}\\}
\address{${}^{(1)}$ Instituto de Ciencias Nucleares \\
 Universidad Nacional Aut\'onoma de M\'exico\\
 Apdo. Postal 70-543, 04510 M\'exico, D.F., MEXICO \\
 $^{(2)}$ Physics Department,  University College Cork \\
Cork, IRELAND \\ }

\maketitle

\begin{abstract}
We establish necessary conditions for the
appearance of both apparent horizons and singularities in
the initial data of spherically symmetric general relativity
when spacetime is foliated extrinsically.
When the dominant energy condition is satisfied these
conditions assume a particularly simple form. Let
$\rho_{\rm Max}$ be the maximum value of the energy
density and $\ell$ the radial measure of its support.
If $\rho_{\rm Max}\ell^2$ is bounded from above by some
numerical constant, the initial data cannot possess an apparent horizon.
This  constant does not depend sensitively on the gauge. An analogous 
inequality is obtained for singularities with some larger constant.
The derivation exploits Poincar\'e type inequalities to bound
integrals over certain spatial scalars. A novel approach to the
construction of analogous necessary conditions for general initial data
is suggested.
\end{abstract}
\date{\today}
\pacs{PACS numbers: 04.20.Cv}

\section{INTRODUCTION}

In this paper we cast necessary conditions for the
appearance of apparent horizons and singularities in the initial data
of spherically symmetric general relativity.
This is the natural sequel to a previous paper in which
we examined sufficiency conditions in the same context \cite{IV}\cite{prev}.
The formulation of necessary conditions is clearly more difficult.
This is because by the nature of the problem, we must assume
the worst: a scenario in which
the geometry possesses an apparent horizon or a singularity.

The initial data consists of the intrinsic and extrinsic geometry on some
spacelike hypersurface. We suppose that the line element on the spatial
geometry is parametrized by

\begin{equation}ds^2= d\ell^2+R^2 d\Omega^2\,.
\label{eq:lineel}\end{equation}
Here $\ell$ is the radial length on the surface. $4\pi R^2$ is
the area of a sphere of fixed $\ell$.
We can express the spherically symmetric
extrinsic curvature in terms
of two spatial scalars, $K_{\cal L}$ and $K_R$ \cite{I}

\begin{equation}
K_{ab}= n_a n_bK_{\cal L} + (g_{ab}-n_a n_b)K_R\,.
\label{eq:scalars}\end{equation}
Here $n^a$ is the outward pointing unit normal to the two-sphere of fixed
radius.
We have that $R$, $K_R$ and $K_{\cal L}$ are constrained by the
hamiltonian and momentum constraints,

\begin{equation}
{1\over 2}(1+ R^{\prime2}) -  (RR')^\prime  =
4\pi\rho R^2 + {1\over 2}R^2 (2K_{\cal L} - K_R)K_R
\,,\label{eq:ham}\end{equation}
and

\begin{equation}K_R^\prime +
{R^\prime \over R}(K_R-K_{\cal L})=4\pi J\,.\label{eq:mom}\end{equation}
The primes represent derivatives with respect to $\ell$.
We assume that both the energy density of the matter
$\rho$ and its current $J$ are finite.

We exploit an extrinsic time foliation.
This involves a constraint on the two
extrinsic curvature scalars.
We will suppose that this constraint is quasi - linear,
and homogeneous so that

\begin{equation}K_{\cal L} + \alpha K_R = 0. \label{eq:alpha}\end{equation}
where $\alpha$ is some  specified not necessarily local
function of the configuration variables,
$\alpha=\alpha(K_R, R, \ell)$ which is
bounded from below by $0.5$ \cite{I,III}.

By a regular geometry, in this paper we understand any spatial geometry
with a single asymptotically flat
region and a regular center, $\ell=0$, without either
apparent horizons or singularities.
The appropriate boundary
condition on the metric at $\ell=0$ is then

\begin{equation}R(0)=0\,.\label{eq:bc}\end{equation}

The occurence of apparent horizons or singularities
in spherically symmetric general relativity
is due entirely to the action of matter.
Though $\rho$ and $J$ are finite there is no guarantee that a
regular asymptotically flat solution defined for all
$\ell\ge 0$ will exist \cite{III}.

At a future (past) apparent horizon, the optical scalar $\omega_{\pm}$
defined by
\cite{MOM}\cite{I,III}

\begin{equation}
\omega_\pm  = 2(R' \pm  RK_R)\label{eq:omega}
\end{equation}
vanishes,

\begin{equation}
\omega_\pm(\ell_H)=0\,.\label{eq:Hbc}
\end{equation}
To avoid clutter we will focus only on future horizons
in this paper.

Singularities occur when the geometry
pinches off at some finite proper radius, $\ell_S$, from the center,

\begin{equation}
R(\ell_S)=0\,.\label{eq:Sbc}
\end{equation}
A singular geometry necessarily contains at least one kind of apparent horizon.
If the mass-function becomes negative as one approaches the singularity one
must have both future and past horizons.

To provide necessary conditions for an
apparent horizon or a singularity we consider the
bounded region enclosed by the feature in question.
The boundary  condition (\ref{eq:Hbc}) or (\ref{eq:Sbc})
is then imposed on Eqs.(\ref{eq:ham}) and
(\ref{eq:mom}).
Integrating Eq.(\ref{eq:ham}) over the domain $[0,\ell_H]$ or
$[0,\ell_S]$ then provides an integrability condition on the
spatial geometry and the sources. This integrability condition then
provides the basis for an
inequality bounding some measure of the energy
content of the region by some measure of its size.

What constitutes a  natural measure of the energy content is a subtle
issue. In our examination of sufficiency conditions
we found that the appropriate measures were the total enclosed
material energy, $M$,
or the difference, $M-P$, where $P$ is the material current
\cite{IV}. We found that
if either the weak or the dominant energy condition holds,
and the geometry did not possess an
apparent horizon, then
$M-P < $ {\it constant} $\ell$, where $\ell$ is
the radial support, for some constant of order unity \cite{IV}.
The same inequality with $M-P$ replaced by $M$ and
with some larger constant is obtained for singularities.

In \cite{II}, where we addressed the problem
when the geometry is momentarily static,
we saw  that $M$ can remain small
though $\ell$ be arbitrarily large. This can occur because
$R$ is folded into the definition of $M$
and $R$ can either saturate or worse become small.
One should not therefore expect $M$ to
serve as a useful measure of the material energy for
the purpose of casting necessary conditions.
Indeed, we know that the statement: if
$M \le$ {\it constant}   $\ell$ then the geometry is
non - singular ---  cannot be justified\cite{weak}.

In \cite{II}, however, with $K_{ab}=0$ we did identify  variables
with respect to which non-trivial necessary conditions could be cast
of the form: if

\begin{equation}\rho_{\rm Max}\ell^2 < {\rm constant}\,,
\label{eq:rholc}\end{equation}
where $\rho_{\rm Max}$ is the maximum energy density,
the distribution of matter will not possess an
apparent horizon with one constant;
with some larger constant it will not
possess a singularity.

When matter flows, the obvious generalization of $\rho_{\rm Max}$
is the sum $\rho_{\rm Max} + J_{\rm Max}$ --- however,
$\rho_{\rm Max}$ and $J_{\rm Max}$ do not
enter symmetrically into the inequality.
Unlike the sufficiency conditions which
involved a symmetric combination of $M$ and $P$,
the equalities we obtain do not respect this symmetry.
The natural inequality we obtain involves not
only $J$ but its square, assuming the form: if

\begin{equation}
\rho_{\rm Max}\ell^2 +
c_0 J_{\rm Max}\ell^2 + c_1(J_{\rm Max}\ell^2)^2<  c_2
\,, \label{eq:rhoJ2lc}\end{equation}
where $c_0$, $c_1$ and $c_2$ are three given numerical constants, then
the geometry is regular.
Once a choice of  gauge has been made the symmetry
between $\rho$ and $J$ is necessarily broken.
Despite appearances this is not an artifact of the
extrinsic time slicing we have exploited.
The value of $J$ plays a more
significant role than the value of $\rho$. This is consistent with
our findings in \cite{III} in our examination of the generic behavior
of the metric in the neighborhood of a singularity in an
$\alpha$-foliation of spacetime.

When the dominant energy condition,

\begin{equation}
\rho \ge |J|\,,\label{eq:DEC}
\end{equation}
is satisfied, the momentarily static form (\ref{eq:rholc}) obtains
from Eq.(\ref{eq:rhoJ2lc}) with some larger constant which depends only
weakly on $\alpha$. This is remarkable in many ways. The single
hamiltonian constraint is replaced by the two coupled
equations, (\ref{eq:ham}) and (\ref{eq:mom}) satisfying
the gauge condition, (\ref{eq:alpha});
in the worst scenario
we must assume,  not only do we need to contend with potential
divergences in the intrinsic geometry but, in addition, with divergences
in the extrinsic curvature.

The paper is organized as follows. In Sect.2, we collect some
relevant bounds on potential divergences.
In Sect.3, we discuss the weights which must
be introduced into  integrals over relevant
geometrical scalars to render them well defined
when the scalar is singular at the end point of the
domain of integration.
In Sect.4,  we derive a necessary condition of the form
(\ref{eq:rhoJ2lc}) for singularities.
In Sect.5, we do the same for apparent horizons.
In Sect.6, we derive a simple necessary condition for the appearance of minimal
surfaces. We end with brief discussion.
Derivations of mathematical inequalities are
provided in an appendix.

\section{Bounds on  $R$, $R'$ and $K_R$}

To formulate a necessary condition for singularities
it is important to possess some bound limiting the
maximum values of $R'$ and $K_R$
which does not require the geometry to be regular.
In particular, one cannot exploit the numerical
bounds on these variables derived in \cite{III} which rely on the
regularity of the geometry. Indeed
these quantities can be arbitrarily large.
What we need to do is place an upper  bound on their rate of
divergence in the neighborhood of singularities.
These bounds will then be applied to determine the
weights which are appropriate to turn the integrability condition into
an inequality.
In fact, this will be their only use in this paper.

We first recall that Eq.(\ref{eq:mom}) can be solved for $K_R$
 in terms of the radial flow of matter, $J$, as follows

\begin{equation}
K_R = {4\pi\over R^{1+\alpha}}
\int_0^\ell d\ell_1\, R^{1+\alpha} J \,\Delta (\ell_1,\ell)
\,.\label{eq:mom1}\end{equation}
The positive factor $\Delta$ is given by

\begin{equation}
\Delta (\ell_1,\ell)= e^{\int_{\ell_1}^\ell d\ell_2 \,\alpha'\ln (R/L)}
\,,
\label{eq:Delta}
\end{equation}
where $L$ is some arbitrary length scale. If $\alpha$ is constant,
$\Delta=1$. This form of the solution makes explicit
the fact that spatial variations of $\alpha$ can
be absorbed into a multiplicative dressing of
the current density.
The constant $\alpha$ result is modulated
by $\Delta$.

It is now straightforward to place a bound on $K_R$.
We have

\begin{equation}
K_R \le  {4\pi |J_{\rm Max}|\over R^{1+\alpha}}
\Delta(0,\ell)
\int_0^{\ell}
d\ell_1\, R^{1+\alpha}
\,.\label{eq:K_Rbound}
\end{equation}
We saw in \cite{III} that $\Delta$
is, in fact, finite everywhere.
It is possible to further bound $\Delta(0,\ell)$ by bounding
$\alpha'$ by
$|\alpha'|_{\rm Max}$ and pulling it through the integral.
However, we will treat the integral appearing in the exponent
itself as the natural measure of the variation of $\alpha$.
Just as we found that $\alpha\ge 0.5$, we will need to bound the
variation of $\alpha$ appropriately if we are not to be overwhelmed by
gauge introduced noise in casting necessary conditions.
Recall that no such bound was ever invoked when we addressed sufficiency
conditions in \cite{IV}.

The exact expression (\ref{eq:mom1}) and the bound (\ref{eq:K_Rbound})
determines the potential divergence of $K_R$ at
a singularity. This occurs with $R$ returning to zero
at some finite radius
from the center, at $\ell = \ell_S$. In \cite{III},
we saw that in the neighborhood of this point

\begin{equation}R\sim
\left({ C_\alpha\over\alpha+1}\right)^{1\over\alpha+1}
(\ell_S-\ell)^{1\over\alpha+1}
\,,\label{eq:Ratsing}\end{equation}
where $C_\alpha$ is the finite constant,

\begin{equation}
C_\alpha=
\int_0^{\ell_S} d\ell_1\, R^{1+\alpha} J \,\Delta (\ell_1,\ell)
\,.
\label{eq:calpha}
\end{equation}
Generically, therefore, $R'$ diverges at $\ell_S$ as do
all higher derivatives of $R$.
If $\alpha(\ell_S)>0.5$, such spatial singularities are more
severe than the strong singularities discussed in \cite{II}
which are consistent with
the Hamiltonian constraint at a moment of time symmetry.
Increasing this value of $\alpha$ increases the strength of
the singularity.

Even if the geometry is singular so that
$R'$ diverges, it can only diverge to minus infinity
--- the surface $R' = 1$  in the configuration space can never be
breached from below. We always have $R'\le 1$\cite{III}.

\section{Poincar\'e Inequalities, Weights and Measures}

Crucial to the derivation of Eq.(\ref{eq:rholc}) in \cite{II} were two
simple Poincar\'e inequalities of the form

\begin{equation}
S\int_0^{\ell_1} d\ell\, \, R^2\le
\int_0^{\ell_1} d\ell\, \, R^{\prime2}
\,,\label{eq:Poinc}\end{equation}
where $S$ depends on the boundary conditions satisfied by $R$.
In general $R(0)=0$. At the first trapped surface, $R'(\ell_1)=0$
and $S= \pi^2/4\ell_1^2$. At a singularity, $R(\ell_1)=0$ and
$S=\pi^2/\ell_1^2$.

Recall that because $R'\le 1$, $R$ is always bounded by $\ell$.
This guarantees that if the
geometry is small in the radial direction it will also be small in the
two transverse directions.
A consequence is that any integral over a positive powers of $R$
will be well defined over any finite interval.
At a singularity, in a moment-of-time-symmetry slice, however, we found that
$R$ tends to zero like $R\sim (\ell_S-\ell)^{2/3}$ so that
$R'$ diverges like $(\ell_S-\ell)^{-1/3}$.  Even though
$R'$ diverges so that the integrand on the RHS of Eq.(\ref{eq:Poinc})
diverges, the integral itself remains finite.
When $J\ne 0$,  $R'$ can diverge more rapidly.
Eq.(\ref{eq:Ratsing}) implies
$R'\sim (\ell_S-\ell)^{-\alpha/1 +\alpha}$.
Thus the integral on the RHS of Eq.(\ref{eq:Poinc}) will only exist if
$\alpha<1$. This is outside the range found to provide the best
sufficiency results in \cite{IV}.
To remedy this situation a non-trivial weight
function will need to be introduced into the integrand
to render the bounding integral well defined.
In \cite{II}, we found that we
could improve the inequalities of necessity at a moment of
time symmetry by weighting with
an appropriate power of $R$. Here it will be essential.

Again, let this function be some power of
$R$, $R^a$ say. The relevant exponent will generally
depend on $\alpha$. At a singularity, $R^a R'^2 \sim
(\ell_S-\ell)^{(a - 2\alpha)/ (1+ \alpha)}$. The
integral

\begin{equation}
\int_0^{\ell_1} d\ell\,R^a R'^2
\end{equation}
will exist for all $a> \alpha -1$. This is not, however, the
optimal value for our purposes. We will see below that
a larger value is desirable.
If $a$ is constant, we have $R^{a/2} R'= (R^{1+a/2})'/ (1 + a/2)$.
We then simply apply Eq.(\ref{eq:Poinc}) to the
function $R^{1 + a/2}$ in place of $R$.

\section{Singularities}

When the gauge condition,
Eq.(\ref{eq:alpha}) is satisfied, we note that
the Hamiltonian constraint
assumes the form

\begin{equation}
{1\over 2}(1+ R^{\prime2}) = (RR')^\prime +
4\pi\rho R^2 + {1\over 2}(2\alpha-1) R^2 K_R^2
\,.\label{eq:ham1}\end{equation}
The second and third terms on the RHS
are manifestly positive. Suppose that the geometry
is singular at $\ell=\ell_S$. We cannot simply integrate Eq.(\ref{eq:ham1})
and discard the boundary term. First of all, as we pointed out above,
it is clear from
Eq.(\ref{eq:Ratsing}) that the integral of $R'^2$ does not
exist on the interval $[0,\ell_S]$;
in addition, the surface
term $RR'$ does not vanish at the singularity unless $\alpha <1$ there.
To remedy the problem we multiply Eq.(\ref{eq:ham1}),
as discussed in Sect.3,
by an appropriate weight function, $R^a$, before integration.

This multiplication has the unfortunate side - effect of destroying the
divergence $(RR')'$ appearing on the RHS of Eq.(\ref{eq:ham1}).
It is, however, simple to  restore this divergence:
we note that

\begin{equation}
( R^{1+ b})' = (1+b) R^{b} R' - b' R^{1+b} \ln R/\ell_S\,.\label{eq:id}
\end{equation}
We perform an integration by parts on the term $R^a(RR')'$, and
now substitute the RHS of Eq.(\ref{eq:id}) for $(R^a)'$ ($a=b+1$):

\begin{eqnarray}
{1\over 2}\int_0^{\ell_S}
d\ell\, R^a(1 + (2a+1) R^{\prime2}) &=& R^{1+a} R^\prime\Big|_{\ell_S}
+ 4\pi \int_0^{\ell_S} d\ell\, \rho R^{2+a} - \Phi\nonumber\\
&& + {1\over 2}
\int_0^{\ell_1} d\ell\,(2\alpha-1)\,R^{2+a} K_R^2
\,,\label{eq:int1}\end{eqnarray}
where $\Phi$, given by

\begin{equation}
\Phi = \int_0^{\ell_S}
d\ell\, a'\ln (R/\ell_S) R^{1+a} R'\,,
\label{eq:Phi}\end{equation}
is a correction which vanishes if $\alpha'=0$.
To discard the boundary term, we require $R^{1+a} R'$ to vanish at
the singularity. This implies that

\begin{equation}a > \alpha-1
\,.\label{eq:a}\end{equation}
This choice of $a$ simultaneously bounds the integral over
$R^a R'^2$.

We also will need to place a bound on the last term on the RHS of
Eq.(\ref{eq:int1}). We exploit Eq.(\ref{eq:K_Rbound}) to bound $K_R$.
The problem is that this bound
involves the positive power of $R$, $R^{1+\alpha}$, in the denominator
which is difficult to control. We obtain the bound,

\begin{equation}
\int_0^{\ell_S} d\ell\, (2\alpha-1)\,R^{2+a} K_R^2\le (4\pi)^2
(2\alpha_{\rm Max}-1) J_{\rm Max}^2
\Delta(0,\ell_S)^2
\int_0^{\ell_S} d\ell\, R^{a - 2\alpha}
\left(\int_0^\ell d\ell\, R^{1+\alpha} \right)^2
\label{eq:intK^2}\end{equation}
on this term. If the weighting exponent is chosen such that

\begin{equation}
a\ge 2\alpha
\,,\label{eq:a1}\end{equation}
the denominator is removed by the weight. Fortunately, such values are
consistent with Eq.(\ref{eq:a}) for all physically acceptable values of
$\alpha$.
The RHS of Eq.(\ref{eq:int1}) is clearly simplest when

\begin{equation}a=2\alpha\,.\label{eq:a2}\end{equation}
This is the value we will henceforth adopt for $a$.

The expression on the right hand side of
Eq.(\ref{eq:intK^2}) is still not very useful as it stands.
A remarkable fact, however, is that we can bound it
by an integral over $R^{2(1+\alpha)}$.
In fact, we have the following inequality

\begin{equation}
\int_0^{\ell_1} d\ell\, \left(\int_0^\ell d\ell\, R^{1+\alpha} \right)^2
\le \left({2\over \pi}\right)^2 \ell_1^2
\int_0^{\ell_1}d\ell\, R^{2(\alpha  +1)}
\,.\label{eq:M1}\end{equation}
This result is derived in the appendix.
Eq.(\ref{eq:M1}) implies the bound for the $K_R^2$ term:

\begin{equation}
\int_0^{\ell_S} d\ell\, (2\alpha-1)\,R^{2(1+\alpha)} K_R^2\le
64 (2\alpha_{\rm Max}-1) J_{\rm Max}^2 \Delta(0,\ell_S)^2
\ell_S^2 \int_0^{\ell_S}d\ell\, R^{2(1+\alpha)}
\,.\label{eq:Kbound}\end{equation}
To understand why this  bound is
important, note that we can exploit
the identity (\ref{eq:id}) to cast
the integrand $R^{2\alpha}R^{\prime2}$ appearing on the
LHS of Eq.(\ref{eq:int1}) in the form

\begin{equation}R^{2\alpha} R^{\prime2}
= {1\over (1+\alpha)^2}\left(
(R^{1+\alpha })'^2 - \alpha'^2 R^{2+2\alpha}
\ln^2 (R/\ell_S)\right)
- {2\alpha'\over 1+\alpha}
R^{1+2\alpha } R' \ln (R/\ell_S)
\,.\label{eq:cast}\end{equation}
If $\alpha$ is constant only the first term survives.
Let us focus on this term.
A one-dimensional Poincar\'e
inequality can be exploited to place a lower bound on the
integral over $(R^{(1+\alpha)})'^2$:

\begin{equation}
S\int_0^{\ell_1}d\ell\, R^{2(1+\alpha)} \le
\int_0^{\ell_1}d\ell\, (R^{1+\alpha})'^2
\,,\label{eq:Poinc1}\end{equation}
where the constant $S =\pi^2 /\ell_1^2$ is the constant
which is relevant for functions which vanish at both $\ell=0$ and
$\ell=\ell_1$.

If $\alpha$ is constant, we then have

\begin{equation}
\int_0^{\ell_S} d\ell\, R^{2\alpha}
 \le  2\left[4\pi \rho_{\rm Max} + 32
(2\alpha-1)  J_{\rm Max}^2 \ell_S^2
-\left({\pi\over \ell_S}\right)^2
{1+4\alpha \over 2(1+\alpha)^2 }\right]
\int_0^{\ell_S} d\ell\, R^{2(1+\alpha)}
\,.\label{eq:ineq1}\end{equation}

In \cite{II}, we proved that when
Eq.(\ref{eq:bc}) is satisfied and
$R'\le 1$ then the ratio of the integrals appearing in
Eq.(\ref{eq:ineq1}) can be bounded as follows (Eq.(6.3.16)) ($a=2\alpha$)

\begin{equation}
{\int_0^{\ell_1} R^{2 + a} d\ell \over
\int_0^{\ell_1} R^{a}d\ell}\, \le {1 + a \over 3 + a}{\ell_1}^2 \,.
\label{eq:ratio}\end{equation}
Eq.(\ref{eq:ratio}) implies

\begin{equation}
{1\over 2} {3 + 2\alpha\over 1 + 2\alpha  }
+ {1+4\alpha \over 2(1+\alpha)^2 }\pi^2
\le 4\pi \rho_{\rm Max}\ell_1^2  +
32 (2\alpha-1) J_{\rm Max}^2 \ell_1^4
\,.\label{eq:ineq2}\end{equation}
We note that it is the second term on the left hand side
which will determine the bound for $\alpha\sim + 1$.
It is maximized when $\alpha= 0.5$.
With this value, we reproduce the
moment of time symmetry result \cite{II}
--- this is a peculiarity of this gauge.

The dependence on the value of $\alpha$ will generally
not be a strong one so long as $\alpha$ is bounded.
In particular, if $\alpha=1$

\begin{equation}{5\pi\over 32}\left[1 + {4\over 3 \pi^2}\right]
\le  \rho_{\rm Max}\ell_1^2 +
{8\over \pi} (J_{\rm Max} \ell_1^2)^2
\,.\label{eq:ineq3}\end{equation}
Note the asymmetry between the roles of $\rho_{\rm Max}$ and $J_{\rm Max}$.
The inequality does not involve what one would to be
the obvious generalization of $\rho_{\rm Max}$,
the sum $\rho_{\rm Max} + J_{\rm Max}$.
$J_{\rm Max}$ plays a more
decisive role than $\rho_{\rm Max}$ in the inequality,
appearing as it does through its square in contrast to
$\rho$ which appears linearly. The inequality with $K_{ab}=0$ does
not generalize in the obvious linear way.
If the dominant energy condition Eq.(\ref{eq:DEC}) holds, the inequality
simplifies. For $\alpha=1$ we obtain

\begin{equation}
{1\over 8}\left[\sqrt{{5\over 3} +{3\over 2}\pi^2}-
{\pi\over 2}\right]
\le  \rho_{\rm Max} \ell_S^2
\,.\label{eq:ineq5}\end{equation}
The LHS $\sim 5/16$, which is approximately half as good as
the moment of time symmetry result.

If $\alpha$ is not a constant, additional noise is introduced
into the inequality by the gauge.
We get

\begin{equation}
\int_0^{\ell_1} d\ell\, R^{2\alpha}
\le  2\left[4\pi \rho_{\rm Max} + 32
(2\alpha_{\rm Max}-1)  J_{\rm Max}^2 \Delta^2 \ell_1^2
-\left({\pi\over \ell_1}\right)^2
{1+4\alpha \over 2(1+\alpha)^2 }\right]
\int_0^{\ell_1} d\ell\, R^{2(1+\alpha)}
+ \Phi_1 + \Phi_2
\,,\label{eq:ineq22}\end{equation}
where $\Phi_1$ and $\Phi_2$ are given
respectively by

\begin{equation}
\Phi_1 = \int_0^{\ell_1} d\ell \, {2\alpha-1\over 1+\alpha}
 \alpha'\ln R/\ell_1 R^{1+2\alpha} R'
\,,
\end{equation}
and

\begin{equation}
\Phi_2 = \int_0^{\ell_1} d\ell \,
{1+4\alpha \over 2(1+\alpha)^2 }
\alpha'^2 \ln^2 R/\ell_1
R^{2(1 + \alpha)}\,.
\end{equation}
The spatial dependence of $\alpha$ is encoded in $\Delta$ and
two terms $\Phi_1, \Phi_2$ which get picked up in the
trade off of $R^{\alpha} R'$ for $(R^{1+\alpha})'$.
 $\Phi_1$ includes the contribution from $\Phi$
appearing in Eq.(\ref{eq:int1}).

These integrals can both
be bounded. We have

\begin{equation}
\Phi_1 \le
{2\alpha_{\rm Max} -1\over 1+\alpha_{\rm Min}}
R_{\rm Max}
^{1+2\alpha} \int_0^{R_{\rm Max}} dR \, |\alpha'||\ln R/\ell_1|
\,,\label{eq:p1}
\end{equation}
and

\begin{equation}
\Phi_2 \le
{1+4\alpha_{\rm Max} \over 2(1+\alpha_{\rm Min})^2 }
R_{\rm Max}^{2(1 + \alpha)} \int_0^{\ell_1} d\ell \, \alpha'^2 \ln^2 R/\ell_1
\,.\label{eq:p2}
\end{equation}
The integrated logarithm appearing in Eq.(\ref{eq:p1}) is bounded by
that which appears in the definition,
Eq.(\ref{eq:Delta}) of $\Delta$. Clearly, we can bound both
by (the square root of) the integral appearing in
Eq.(\ref{eq:p2}). This is the only measure of
$\alpha'$ we need to control.
We will also need the bounds

\begin{equation}
R_{\rm Max}^{n + 2\alpha} \Big/
\int_0^{\ell_1} d\ell\, R^{2\alpha}
\le (1+2\alpha_{\rm Max}) \,\ell_1^{n-1}\,,
\label{eq:ratio1}\end{equation}
for $n\ge 1$.

\section{Apparent Horizons}

At a moment of time symmetry, there is a remarkable similarity
between the signal for the presence of an apparent horizon,
$R^\prime=0$  and that for the presence of a singularity,
$R=0$. In \cite{II}, this meant that the techniques we exploited
for analysing singularities were also good for
analyzing apparent horizons and the effort required almost identical.
In general, however, the signal for an apparent horizon
will involve the extrinsic curvature of the spatial hypersurface
through Eq.(\ref{eq:Hbc}).
Its physical location no longer coincides with an extremal
surface of the spatial geometry as it did at a
moment of time symmetry.

At a future apparent horizon, $\omega_+$ defined by
Eq.(\ref{eq:omega}) vanishes.
Eliminating $R'$ in the divergence term in Eq.(\ref{eq:ham})
using Eq.(\ref{eq:omega}) we obtain

\begin{equation}
{1\over 2}(1+ R^{\prime2}) = (R\omega_+ - R^2 K_R)^\prime +
4\pi\rho R^2 + {1\over 2}(2\alpha-1) R^2 K_R^2
\,.\label{eq:hor1}\end{equation}
Again both the second and third terms on the RHS are manifestly positive.
Let us suppose for simplicity that $\alpha$ is constant.

Suppose that all quantities are well
defined (we will relax this assumption below).
We can then integrate Eq.(\ref{eq:hor1}) up to the first future horizon
at which $\omega_+=0$ to get

\begin{equation}
\int_0^{\ell_1} d\ell\, (1 + R'^2)=
- R^2 K_R\Big|_{\ell_1}+  4\pi \int_0^{\ell_1} d\ell\, R^2 \rho +
{1\over 2} (2\alpha-1) \int_0^{\ell_1} d\ell\,
R^2 K_R^2
\,.\label{eq:inthor1}\end{equation}
We wish to exploit Eq.(\ref{eq:K_Rbound})
to place a bound on $K_R$ in the surface term.
Unfortunately, this bound will only be well
defined for $\alpha\le 1$.

The first two terms can be dealt with symmetrically when $\alpha=1$. In this
case these first two terms on the  RHS can be bounded as follows:

\begin{equation}
- R^2 K_R\Big|_{\ell_1}+  4\pi \int_0^{\ell_1} d\ell\, R^2 \rho
\le
4\pi \Big(\rho_{Max} + |J_{Max}|\Big)\int_0^{\ell_1}
d\ell\, R^2
\,.\label{eq:G1}\end{equation}
A linear term in $J_{\rm Max}$ appears in the apparent
horizon inequality condition which is not present in the
singularity inequality. This is a reflection of the
different boundary conditions enforced there.

We can exploit a Poincar\'e inequality to place a bound on the integral
over the interval $[0,\ell_1]$ of the quadratic $R^2$ by the same
integral over the quadratic, $R^{\prime2}$:

\begin{equation}
S\int_0^{\ell_1} d\ell \, R^2\le
\int_0^{\ell_1} d\ell \, R^{\prime2}
\,.\label{eq:P1}\end{equation}
The inequality is saturated by the trigonometric
function,

\begin{equation}
R(\ell) = \sin (\gamma \ell)
\,,\label{eq:sin}\end{equation}
which also determines the optimal value of $S=\gamma^2$. The boundary
condition, (\ref{eq:Hbc}) determines $\gamma$ to be the lowest solution of
the transcendental equation,

\begin{equation}
\tan \gamma \ell_1 = -{\gamma\over K_R}
\,.\label{eq:tan}\end{equation}
We note that
\begin{equation}
\gamma \le {\pi\over 2\ell_1}
\label{eq:gamma}\end{equation}
if $K_R$ is negative with $\gamma \to \pi/2\ell_1$ as $K_R\to 0$
which
is the moment of time
symmetry bound and $\gamma \to \pi/\ell_1$ as $K_R\to +\infty$.

Unfortunately, even when $\alpha = 1$, when we attempt to bound the
third term on the right hand side  we run into the same problem we
faced when we examined singularities in Section 4 with the same term.
We need to introduce a weighting to guarantee convergence of the integral. The
same weighting which worked for singularities  works again. There is no
real simplification in the $\alpha = 1$ case so we will return to the general
case. To restore the divergence appearing in Eq.(\ref{eq:hor1}) we need to
perform an integration by parts as before. We integrate up to
$\ell_1$:

\begin{eqnarray}
{1\over 2}\int_0^{\ell_1}
d\ell\, R^{2\alpha}(1 + (4\alpha+1) R^{\prime2})  &= &
-R^{2(1+\alpha)} K_R\Big|_{\ell_1}
+ 4\pi \int_0^{\ell_1} d\ell\, \rho R^{2(1+\alpha)}\nonumber\\
&& + {1\over 2}(2\alpha-1) \int_0^{\ell_1} R^{2(1+\alpha)} K_R^2
\,.\label{eq:inthor2}\end{eqnarray}
We now exploit Eq.(\ref{eq:K_Rbound}) to bound the $K_R$ and $K_R^2$ terms.
For the former,

\begin{equation}
R^{2(1+\alpha)} K_R\Big|_{\ell_1} \le
4\pi R^{1+\alpha} J_{\rm Max} \int_0^{\ell_1} d\ell\,
R^{1+\alpha} \,.\label{eq:5.39}\end{equation}
The weighting process has broken the symmetry under interchange
of $\rho$ and $J$ of the linear terms on the RHS of Eq.(\ref{eq:inthor1})
which is evident in Eq.(\ref{eq:G1}).
For the term quadratic in $K_R$, we again have ((\ref{eq:Kbound}) with
$a=2\alpha$)

\begin{equation}
\int_0^{\ell_1} d\ell\,R^{2(1+\alpha)} K_R^2\le (4\pi)^2  J_{\rm Max}^2
\int_0^{\ell_1} d\ell\,\left(\int_0^\ell d\ell\, R^{1 +\alpha} \right)^2
\,.
\end{equation}
We again require a bound on the last term by an integral over
$R^{2(1+\alpha)}$. Though the boundary
conditions are different we again obtain the bound (\ref{eq:M1}).
We demonstrate this in the appendix.
We can now write

\begin{eqnarray}
1 \le 2\left[4\pi \rho_{\rm Max} +
32(2\alpha-1) J_{\rm Max}^2 \ell_1^2
-\tilde{\gamma}^2
{1+4\alpha \over 2(1+\alpha)^2 }\right]
&\int_0^{\ell_1} d\ell\, R^{2(1+\alpha)}\Big/
\int_0^{\ell_1} d\ell\, R^{2\alpha} \nonumber\\
+8\pi J_{\rm Max}&R_1^{1+\alpha}
\int_0^{\ell_1} d\ell\, R^{1+\alpha} \Big/
\int_0^{\ell_1} d\ell\, R^{2\alpha} \,.
\label{eq:ineq6}\end{eqnarray}
Here $\tilde{\gamma}$ is the analogue of the $\gamma$ that appears in
Eqs.(\ref{eq:sin} - \ref{eq:tan}), except that $R^2$ in Eq.(\ref{eq:P1}) is
replaced by  $R^{2(1+\alpha)}$. This means that Eq.(\ref{eq:tan}) must be
replaced by
\begin{equation}
\tan \tilde{\gamma} \ell_1 = -{\tilde{\gamma}\over K_R(1 + \alpha)}
\,.\label{eq:tan1}\end{equation}
The same upper and lower bounds on
$\tilde{\gamma}$ hold,
 {\it i.e.}, $\pi/2\ell \le \tilde{\gamma} \le \pi/\ell_1$.
We can again exploit (\ref{eq:ratio}) to bound the ratio of the integrals in
the
first term of (\ref{eq:ineq6}). In the second term,  one can exploit

\begin{equation}
R_1^{1+\alpha}
\int_0^{\ell_1} d\ell\, R^{1+\alpha} \Big/
\int_0^{\ell_1} d\ell\, R^{2\alpha}
\le {1+2\alpha\over 2+\alpha}\,\ell_1^2\,.
\label{eq:ratio2}\end{equation}
This is proved using the same
technique as the derivation of Eq.(\ref{eq:ratio}).
The necessary
condition for an apparent horizon with constant
$\alpha$ is then

\begin{equation}
4\pi \left(\rho_{\rm Max} +
{3+2\alpha\over 2 + \alpha} J_{\rm Max} \right)\ell_1^2
+32 (2\alpha-1) J_{\rm Max}^2 \ell_1^4\le {1\over 2}
{3+2\alpha\over 1+2\alpha}
+ {1+4\alpha \over 8(1+\alpha)^2 }\pi^2
 \,.\label{eq:ineq7}\end{equation}
The only real difference with respect to
Eq.(\ref{eq:ineq2}) is the appearance of the linear $J_{\rm max}$
term. When the dominant energy condition is satisfied,
we can replace $J_{\rm Max}$ with $\rho_{\rm Max}$ and get a quadratic
expression in $\rho_{\rm Max}\ell_1^2$. This in turn can be solved to give a
direct bound on $\rho_{\rm Max}\ell_1^2$. When  $\alpha=1$, this becomes
\begin{equation}
 \rho_{\rm Max}\ell_1^2 \le {1 \over 8}\sqrt{{301\pi^2 \over 144} +
{5 \over 3}} - {\pi \over 6} \approx 0.07\,.
\end{equation}
This is approximately three times smaller than the constant we obtained for the
moment of time symmetry case in \cite{II}.

\section{Minimal Surfaces}

There is a very simple necessary condition for the existence of a minimal
surface which is easy to derive and which is essentially gauge independent..
Let us return to the Hamiltonian constraint, Eq.(\ref{eq:ham}). This can be
rewritten as
\begin{equation}
{1\over 2}(1+ R^{\prime2}) -  (RR')^\prime  = {1 \over 4}R^2\,^3{\cal R}\,,
\label{eq:ham1*}\end{equation}
where $^3{\cal R}$ is the three scalar curvature of the initial slice. If
the weak energy condition is satisfied and if $0.5 \le \alpha < \infty$ we have
that
$^3{\cal R}\ge 0$. This is sufficient to show that $R^{\prime} \le 1$.
Let us assume that the initial data contains a minimal surface and that the
first minimal surface occurs at $\ell = \ell_M$. Clearly, in the range
$0 \le \ell \le \ell_M$, we have $0 \le R^{\prime} \le 1$. Let us integrate
Eq.(\ref{eq:ham1*}) from the origin out to $\ell_M$. We get
\begin{equation}
\ell_M \ge {1\over 2}\int_0^{\ell_M}(1+ R^{\prime2})d\ell   =
{1 \over 16\pi}\int_0^{\ell_M}4\pi R^2\, ^3{\cal R}d\ell
= {1 \over 16\pi}\int_0^{\ell_M} \,^3{\cal R} dv\,.
\label{eq:ham2}\end{equation}
The boundary term can be discarded because $R^{\prime} = 0$ at a minimal
surface. Thus a necessary condition for the appearance of a minimal surface is
\begin{equation}
16\pi \ell \ge \int_0^{\ell_M} R^{(3)} dv\,.
\label{eq:ham3}\end{equation}
If we have a minimal surface it must be either future or past trapped.
Unfortunately, we cannot use this condition, Eq.(\ref{eq:ham3}), to derive a
necessary condition for trapped surfaces because we could have a trapped
surface without any minimal surface.

\section{Conclusions}

In this paper we have presented
new necessary conditions for the
presence of both apparent horizons and singularities in
spherically symmetric initial data.

While we have assumed that spacetime is foliated
extrinsically, this is not a
severe restriction. Indeed, modulo the constraints,
the destinction between intrinsic and extrinsic foliations
becomes an artificial one.

The inequalities do not depend sensitively on $\alpha$.
We have seen that just as one has to place a
lower bound on $\alpha$ to obtain a sensible gauge,
to obtain necessary conditions one
needs also to impose an explicit upper bound on the
spatial variation of $\alpha$.
Acting as it does to mask the underlying physics, it is not
at all surprising
that the variation of $\alpha'$ needs to bounded.
It is, overall,  surprising that
all of the gauge ambiguity can be absorbed
in such a simple way.

Our approach to functional analysis has been extremely
heuristic  --- it is clear that some of the inequalities
exploited in Sect. 4 and 5  can be sharpened,
specially those relating to non-constant $\alpha$. As physicists,
however, we always use the
gauge which makes life easiest --- linear gauges with $\alpha$
constant does this. When $\alpha$ is
not constant,  we are clearly more interested in the
fact that such bounds can be established than in squeezing them
for better constants.

How is this work likely to be generalized?
The obvious challenge is to generalize it
to non-spherically symmetric geometries.

The Hoop conjecture
formulated many years ago by Kip Thorne\cite{Th} states, in
rough terms, that a black hole hole will form if and only if
energy is compressed in all three spatial directions.
If we admit `cosmic censorship' the conjecture
can be rephrased in terms of initial data,
with black hole replaced by apparent horizon.
It should be clear why the phrasing of the conjecture is
vague. Even with no independent gravitational degrees of freedom to
worry about, it is remarkably difficult
to provide a description
of the two ingredients `quantity of matter' and size which is simultaneously
valid for both necessity and sufficiency,
never mind proving the conjecture.
The situation can only get worse
when we relax spherical symmetry.
One needs to bear in
mind that our ability to describe the configuration
space in considerable detail has
relied on features of the spherically symmetric problem which, we know,
do not admit generalizations.
Progress has been made on the sufficiency part of the conjecture
\cite{Hoop}. Much less is known about the
necessary part. Our work in this paper
where the Poincar\'e ineqality on the interval plays a central
role, suggests a new approach to attacking the problem in
non-spherically symmetric geometries. This generalization
might involve a Sobolev  type inequality
on the scale factor, $\Phi$:

\begin{equation}
S \left(\int d^3x\, \Phi^6\right)^{1/3} \,\le\,
\int d^3x\,(\nabla\Phi)^2\,.
\end{equation}
Indeed,  had we exploited
conformal coordinates, with respect to which
the spatial line element assumes the form,
$ds^2=\Phi^4 ds_{\rm Flat}^2$, we would have found
ourselves in need of such an inequality to derive the results
of this paper.

We are encouraged by the fact that Sobelev inequalities
are known to be related intimately with the isoperimetric
problem \cite{Chavel}.

A physically interesting question that is extremely relevant is the
identification of initial data that potentially might develop
apparent horizons. In principle it should
be possible to do this exploiting in addition to the
constraints, the dynamical Einstein equations
evaluated on the initial hypersurface. These equations involve the
pressure of matter though some equation of state.
The scenario which is most susceptible
to collapse is pressureless matter. We should be able to
exploit this condition to formulate necessary conditions.
At the other extreme, a stiff equation of
state would inhibit collapse. Thus such a scenario might provide
a sufficient condition.

\section*{Acknowledgements}
We gratefully acknowledge support from CONACyT
Grant 211085-5-0118PE to JG and Forbairt Grant SC/96/750 to N\'OM.

\section*{Appendix}
In this appendix, we provide a derivation of the bound
for the the extrinsic curvature quadratic
used in the text.

\begin{equation}
\int_0^{\ell_1} d\ell\, \left(\int_0^\ell d\ell\, R^{1+\alpha} \right)^2
\le \left({2\over \pi}\right)^2 \ell_1^2
\int_0^{\ell_1}d\ell\, R^{2(\alpha  +1)}
\,.\label{eq:A1}\end{equation}

The existence of a bound of this form
is not hard to see. A crude bound is provided by the
positivity of the covariance for any power $R^n$:(H\"older
Inequality),

\begin{equation}
<R^n>^2\quad \le\quad <R^{2n}>\,,\label{eq:covar}\end{equation}
which implies

\begin{equation}
\left(\int_0^{\ell_1} d\ell\, R^n\right)^2\le
\ell_1 \int_0^{\ell_1} d\ell\, R^{2n}
\,,\label{eq:covar1}\end{equation}
so that
\begin{equation}\int_0^{\ell_1} d\ell\, \left(\int_0^\ell d\ell\, R^{1+\alpha}
\right)^2
\le
{\ell_1^2\over 2} \int_0^{\ell_1} d\ell \,R^{2(1+\alpha)}
\,.\label{eq:covar2}\end{equation}
The bound (\ref{eq:A1}) is, however, better.
To derive it, let

\begin{equation}
G(\ell) :=    \int_0^\ell d\ell\, R^n
\,.\label{eq:G}\end{equation}
Now $G(0)$ =0 and $G'(\ell_1) =0$, for all $n\ge 0$. We apply the
Poincar\'e inequality to $G$ with the constant
which is appropriate with these boundary conditions:

\begin{equation}
\int_0^{\ell_1} d\ell \, G_n(\ell)^2 \le
\left({2\ell_1\over \pi}\right)^2
\int_0^{\ell_1} d\ell \, R^{2n}\,.\label{eq:PoincG}\end{equation}
so that

\begin{equation}
\int_0^{\ell_1} d\ell \,\left(\int_0^{\ell}
d\ell\, R^{(1+\alpha)}\right)^2
\le\left({2\ell_1\over \pi}\right)^2
\int_0^{\ell_1} d\ell \, R^{2(1+\alpha)}
\,.\label{eq:PoincG1}\end{equation}
This is better by a factor of $\pi^2/8$ than the estimate (\ref{eq:covar2}).

The same bound is obtained for functions $R(\ell)$ satisfying
Eq.(\ref{eq:Hbc}) at $\ell=\ell_1$.
The crude bound we derived before, (\ref{eq:covar2}), is
expected to work better this time.
As before, however, we can do better. This time we let

\begin{equation}
H(\ell) :=    \int_0^\ell d\ell\, R^n \Big/\int_0^{\ell_1} d\ell\, R^n
\,.\label{eq:H}\end{equation}
Now $H(0)$ =0 and $H(\ell_1) =1$ for all $n$. We apply the Poincar\'e
inequality to $H$ with the appropriate constant

\begin{equation}
\int_0^{\ell_1} d\ell \, H(\ell)^2 \le {1\over \gamma^2}
\left({2\ell_1\over \pi}\right)^2
\int_0^{\ell_1} d\ell \, R^{2n}
\Big/\Big(\int_0^{\ell_1} d\ell\, R^n\Big)^2 \,,\label{eq:PoincH1}
\end{equation}
where $\gamma$ is given by Eq.(\ref{eq:gamma}).
Exploiting the lower bound on $\gamma$ obtained in the text
we obtain Eq.(\ref{eq:A1}) exactly as we did for singularities.

\vfill
\eject


\begin{references}
\bibitem{weak}
In \cite{II}, in our examination of a moment of time
symmetry, a weaker statement was formulated: if
$M <\ell/2$ on a given surface then that surface is not trapped.
If $J\ne 0$, however, even a statement of this form is beyond our reach.
The reader is referred to \cite{II} for details.
\end{references}
\end{document}